\documentclass[letterpaper, 10pt, conference]{ieeeconf}  

\IEEEoverridecommandlockouts           
\usepackage{epsfig}			
\usepackage{graphicx}		
\usepackage{subfigure}		
\usepackage{url}			
\usepackage{listings}       

\title{\LARGE \bf Paving the Way for Culturally Competent Robots: a Position Paper} 
\author{
	Barbara Bruno$^{1}$, Nak Young Chong$^{2}$, Hiroko Kamide$^{3}$, Sanjeev Kanoria$^{4}$,\protect\\Jaeryoung Lee$^{5}$, Yuto Lim$^{2}$, Amit Kumar Pandey$^{6}$, Chris Papadopoulos$^{7}$,\protect\\Irena Papadopoulos$^{8}$, Federico Pecora$^{9}$, Alessandro Saffiotti$^{9}$ and Antonio Sgorbissa$^{1}$%
	\thanks{ $^{1}$ B. Bruno and A. Sgorbissa are with the University of Genova, Via Opera Pia 13, 16145 Genova, Italy.}
	\thanks{ $^{2}$ Y. Lim and N.Y. Chong are with the Japan Advanced Institute of Science and Technology, 1-1 Asahidai, Nomi, Ishikawa 923-1292, Japan.}
	\thanks{ $^{3}$ H. Kamide is with Nagoya University, Furocho, Chikusaku, Nagoya, Aichi 464-8601, Japan.}
	\thanks{ $^{4}$ S. Kanoria is with Advinia Health Care Limited LTD, Regents Park Road 314, London N3 2JX, United Kingdom.}
	\thanks{ $^{5}$ J. Lee is with Chubu University, 1200 Matsumoto-cho, Kasugai, Aichi 487-8501, Japan.}
	\thanks{ $^{6}$ A.K. Pandey is with Softbank Robotics Europe SAS, Rue Colonel Pierre Avia 43, 75015 Paris, France.}
	\thanks{ $^{7}$ C. Papadopoulos is with the University of Bedfordshire, Park Square, Luton LU1 3JU, United Kingdom.}
	\thanks{ $^{8}$ I. Papadopoulos is with Middlesex University Higher Education Corporation, The Burroughs, Hendon, London NW4 4BT, United Kingdom.}
	\thanks{ $^{9}$ F. Pecora and A. Saffiotti are with \"Orebro University, Fakultetsgatan 1, S-70182 \"Orebro, Sweden.
	\protect\\{Corresponding author's email: \tt\small antonio.sgorbissa@unige.it}}%
}

\begin{document}
\begin{titlepage}  

\copyright 2017 IEEE Personal use of this material is permitted. Permission from IEEE must be obtained for all other uses, in any current or future media, including reprinting/republishing this material for advertising or promotional purposes, creating new collective works, for resale or redistribution to servers or lists, or reuse of any copyrighted component of this work in other works.

\end{titlepage}
\maketitle
\thispagestyle{empty}
\pagestyle{empty}

\begin{abstract}

Cultural competence is a well known requirement for an effective
healthcare, widely investigated in the nursing literature. We claim that
personal assistive robots should likewise be culturally competent, aware
of general cultural characteristics and of the different forms they take
in different individuals, and sensitive to cultural differences while
perceiving, reasoning, and acting. Drawing inspiration from existing
guidelines for culturally competent healthcare and the state-of-the-art
in culturally competent robotics, we identify the key robot capabilities
which enable culturally competent behaviours and discuss methodologies
for their development and evaluation.

\end{abstract}

\section{Introduction}
\label{sec:introduction}

Designers of social and assistive robots are often faced with questions
such as: ``How should the robot greet a person?'', ``Should the robot
avoid or encourage physical contact?'', ``Is there any area of the house
that it should consider off-limits?''.  Intuitively, the correct answer
to all those questions is ``It depends'', and more precisely, it depends
on the person's values, beliefs, customs and lifestyle, i.e., the
person's own \emph{cultural identity}.

The need for cultural competence in healthcare has been widely
investigated in the nursing literature~\cite{Leininger02}.  The fields
of Transcultural Nursing and Culturally Competent Healthcare play a
crucial role in providing culturally appropriate nursing care, as the
presence of dedicated cultural competence international journals and
worldwide associations reflects~\cite{ETNA_online,CARE_online}.
In spite of its crucial importance, cultural competence has been almost
totally neglected by researchers and developers in the area of assistive
robotics.  Today it is technically conceivable to build robots
---possibly operating within a smart ICT environment
\cite{Wongpatikaseree12}--- that reliably accomplish basic assistive
services.  However, these robots only address the problem of ``what to
do'' to provide a service, and produce rigid recipes which are invariant
with respect to the place, person and culture.  We argue that this is
not sufficient and necessarily doomed to fail: if service robots are to
be accepted in the real world by real people, they must take into
account the cultural identity of their users in deciding ``how'' to
provide their services.

\begin{figure}
\centering
\includegraphics[width=8.6cm]{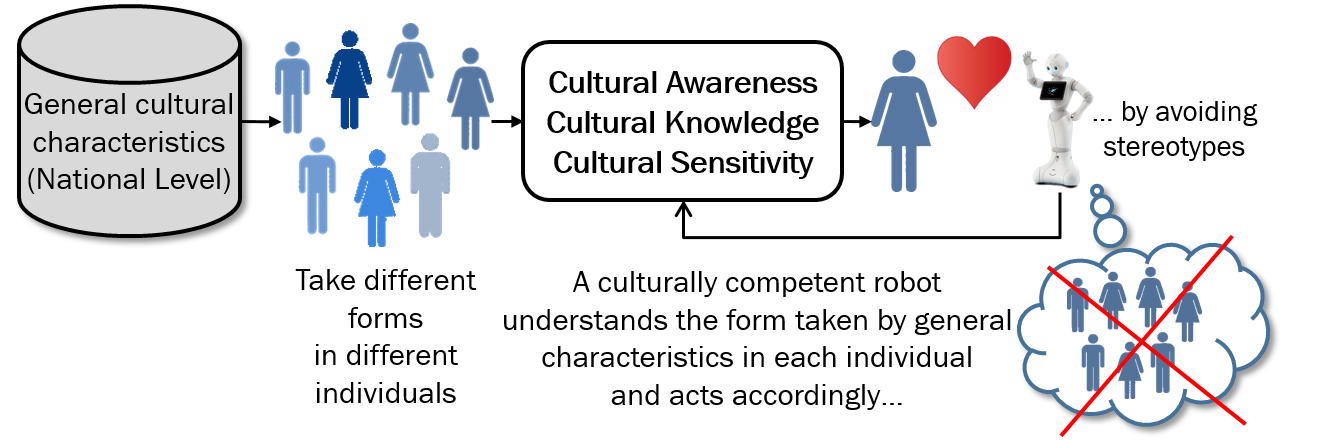}
\caption{The rationale for a culturally competent robot.}
\label{fig:idea}
\end{figure}

In this position paper we discuss the concept of a \emph{culturally
  competent robot}.  We claim that cultural competence is a key factor
for social, personal and assistive robots, which has been mostly
neglected so far.
Figure \ref{fig:idea} illustrates the concept of a culturally competent
robot.  Such a robot (i) knows general cultural characteristics,
intuitively, those characters that are shared by a group of people; (ii)
it is aware that general characteristics take different forms in
different individuals, thus avoiding stereotypes; and (iii) it is
sensitive to cultural differences while perceiving, reasoning, and
acting.  These robots will be able to adapt how they behave and speak to
the culture, customs and manners of the person they interact with.  We
believe that cultural competence is especially important in assistive
robots, where it can increases their acceptability and effectiveness,
which will help to improve the quality of life of users and their
caregivers, support active and healthy ageing, and reduce caregiver
burden.

The contribution of this article is three-fold: (i) to provide a precise
definition of cultural competence in robotics, together with an overview
of current efforts in culturally competent robots; (ii) to analyse the
capabilities which are needed to enable culturally competent robot
behaviour; and (iii) to propose a concrete methodology for their
development and evaluation.  The methodology is grounded in the work
currently performed in the \textit{CARESSES} project, a joint EU-Japan
effort started in 2017 which aims to develop and evaluate a culturally
competent robot for elderly assistance (see
\url{www.caressesrobot.org}).
%

The next sections introduce the three contributions mentioned above,
followed by a discussion on the testing and evaluation methodology and
by some concluding remarks.


\section{Robots and Cultural Competence}
\label{sec:robots_and_culture}

In order to make the concept of a culturally competent robot precise
enough to discuss its technical requirements, we first introduce several
notions related to cultural competence in general.

\subsection{Facets of cultural competence}
\label{subsec:facets_cultural_competence}

Culture and cultural competence are difficult terms to define.  In our
work, we adopt the following definitions taken from the field of
transcultural health and social care~\cite{Papadopoulos06}.

\textbf{Culture}. All human beings are cultural beings. Culture is the
shared way of life of a group of people that includes beliefs, values,
ideas, language, communication, norms and visibly expressed forms such
as customs, art, music, clothing, food, and etiquette. Culture
influences individuals' lifestyles, personal identity and their
relationship with others both within and outside their culture. Cultures
are dynamic and ever changing as individuals are influenced by, and
influence their culture, by different degrees. 

\textbf{Cultural identity}. The concept of identity refers to an image
with which one associates and projects oneself. Cultural identity is
important for people's sense of self and how they  relate to
others. When a nation has a cultural identity it does not mean that it
is uniform. Identifying with a particular culture gives people feelings
of belonging and security. 

\textbf{Cultural awareness}. Cultural awareness is the degree of
awareness we have about our own cultural background and cultural
identity. This helps us to understand the importance of our cultural
heritage and that of others, and makes us appreciate the dangers of
ethnocentricity. Cultural awareness is the first step to developing
cultural competence and must therefore be supplemented by cultural
knowledge. 

\textbf{Cultural knowledge}. Meaningful contact with people from
different ethnic groups can enhance knowledge around their health
beliefs and behaviours as well as raise understanding around the
problems they face. 

\textbf{Cultural sensitivity}. Cultural sensitivity entails the crucial
development of appropriate interpersonal relationships. Relationships
involve trust, acceptance, compassion and respect as well as
facilitation and negotiation. 

\textbf{Cultural competence}. Cultural competence is the capacity to
provide effective care taking into consideration people's cultural
beliefs, behaviours and needs. It is the result of knowledge and skills
which we acquire during our personal and professional lives and to which
we are constantly adding. The achievement of cultural competence
requires the synthesis of previously gained awareness, knowledge and
sensitivity, and its application in the assessment of clients' needs,
clinical diagnosis and other caring skills. 

\begin{table*}[t]
\caption{Introduction scenario: Mrs Christou, a 75 years old Greek
  Cypriot who migrated to the UK when she was 20 years old.} 
\label{tab:Introduction}
\begin{center}
\begin{tabular}{|p{6.8cm}|p{3.8cm}|p{5.9cm}|}
\hline
Scenario & Robot skills & Cultural competence \\
\hline
ROBOT: Hello Mrs. Christou! & Perception (Face recognition) & \\
\textit{The robot hugs Mrs. Christou} & Moving (Arms) & \\
MRS CHRISTOU: Hello! & & \\
\textit{Mrs. Christou smiles and hugs the robot} & & \\
ROBOT: Would you prefer me to call you Kyria Maria? & Speaking (Asking for yes/no confirmation) & [Cultural Knowledge: general (1)] The Greek Cypriot culture is very similar to that of Greece, in which hierarchy should be respected and some inequalities are to be expected and accepted.\\
MRS CHRISTOU: Yes, that's how one calls an older woman in Cyprus. What is your name? &  & [Cultural Awareness (2)] Mrs. Christou values her culture and its customs. She expects others to treat her older age status with some respect: this is why she likes that the robot calls her Kyria Maria (Kyria is Greek for Mrs).\\
ROBOT: I don't have a name yet. Would you like to give me a name? & Speaking (Catching key words and reacting) & \\
\textit{The robot leans slightly forward} & Moving (Body posture) & \\
KYRIA MARIA: I will call you Sofia after my mother, God rest her soul. & & [Cultural Awareness (3)] She names the robot after her mother, a common custom to name one's children. She shows her respect to the dead through signs of her religiosity. \\
\textit{The robot asks for confirmation for the name, infers that Sofia is the name of Kyria Maria's mother and asks for confirmation} & Speaking (Catching key words; asking for yes/no confirmation) & \\
ROBOT SOFIA: Thank you, I like the name. I am honoured to be called after your mother. & &\\
\textit{The robot smiles and hugs Kyria Maria} & Moving (Arms) & \\
\hline
\end{tabular}
\end{center}
\end{table*}

\subsection{Cultural competence in robotics until today}
\label{subsec:background}

An analysis of the literature on personal, social and assistive robots
reveals that the issue of cultural competence has been largely
under-addressed, and a lot of work is still to be done to pave the way
to culturally competent robots.


It is an established fact that robots are generally treated by people as
social actors and expected to comply with social norms~\cite{Evers08}.
As an example, two different studies on the interpersonal distance
between a person and a robot report that people (i) mostly conform to
Hall's social zones when approaching a robot, thus acknowledging it as a
social actor~\cite{Walters05} and (ii) prefer a robot that stays out of
people's intimate space zone, thus expecting it to behave as a
socially-competent actor~\cite{Joosse14b}. A study on the acceptability
of a robot navigating a human environment found that a robot programmed
to respect four basic social conventions was preferred over one lacking
this knowledge~\cite{Pandey09}.

A number of studies support the hypothesis that people from different
cultures not only (i) have different preferences concerning how the
robot should be and behave~\cite{Evers08}, but also (ii) tend to prefer
robots better complying with the social norms of their own culture, both
in the verbal~\cite{Wang10,Andrist15} and non-verbal behaviour
\cite{Eresha13,Joosse14b}.  This preference does not merely affect the
robot's likeability.  In a series of experiments on the influence of
culture on Human-Robot Interaction, participants from the USA and China
were asked to solve a task with the possibility of using the suggestions
of a robot assistant~\cite{Evers08}.  Experimenters analysed the level
of trust, comfort, compliance, sense of control and anthropomorphism
inspired by the robot and found that participants had more trust and a
more effective interaction with the robot complying with the norms of
their own culture~\cite{Wang10}.

Lastly, an innovative take on the analysis of the effect of culture on
the interaction between a person and a robot investigates whether
cultural similarities entail similar preferences in the robot's
behaviour.  An experiment with Dutch participants and two robots,
respectively customized for the German and Japanese culture, provides
preliminary support to the hypothesis that acceptance of a robot could
be directly proportional to cultural closeness~\cite{Trovato15}.


Despite the aforementioned findings, little work has been reported on
how to build robots that can be easily adapted to a given cultural
identity.

Torta et al.~\cite{Torta11} propose a method to parametrize the
interpersonal distance and direction of approach that the robot should
use when talking to a person.  They first define a function (Region of
Approach) with higher values for distances and orientations which are
found to be comfortable for the user and lower values for other
distances and orientations. Then, they combine this function with path
planning information in a Bayesian inference mechanism to identify a
suitable target pose for the robot.

A complex example of cultural adaptation explores a framework for the
learning and selection of culturally appropriate greeting gestures and
words~\cite{Trovato15b}. 
%
%
In the proposed architecture, an initial set of gestures and words is
extracted from video and text corpora, and initial associations between
gestures and words and cultural factors are drawn from literature in
social studies and expressed as conditional probabilities in a Naive
Bayes classifier.  At run-time, the user's cultural background, stored
as a vector of cultural factors, is used to identify the greeting
gestures and words which better match his/her profile.  A
post-interaction questionnaire is then used as a feedback for the
classifier, to allow for an on-line update of the association between
cultural profiles and greeting gestures and words.

Both of the reported works consider adaptation at a personal level, and
follow a ``bottom-up'' approach, i.e. they identify nations as clusters
of people with similar cultural profiles.  The major limitation of this
approach is that it is not well suited for encoding cultural information
expressed at national-level, nor how such information influences
preferences in the robot behaviours. As such, adaptation to a different
culture is a demanding process which requires either a long time, or a
large corpus of data to begin with.

A first attempt at developing a ``top-down'' approach explores the use
of national-level cultural information for the cultural customization of
the gestures and facial expressions of a virtual agent~\cite{Rehm07}.
Among the most popular metrics for the description of culture at a
national-level, Hofstede's dimensions for the cultural categorization of
countries are six scales in which the relative positions of different
countries are expressed as a score from 0 to 100~\cite{Hofstede91}. As
an example, the dimension of \textit{Individualism} examines whether a
nation has a preference for a loosely-knit social framework, in which
individuals are expected to take care of only themselves and their
immediate families, or for a tightly-knit framework, in which
individuals can expect their relatives or members of a particular
in-group to look after them, a notion which Hofstede called
\textit{Collectivism}.

More recently, Hofstede's dimensions have been used to express the
influence of culture on the gestures and words that a robot should use
at a first meeting with a person~\cite{Lugrin15}. The proposed framework
is among the very first attempt at merging the ``top-down'' and
``bottom-up'' approaches, since the system makes use of empirical data
(a corpus of tagged video recordings of pairs of people from the same
country meeting for the first time) to complement the theoretical values
given by Hofstede's dimensions.



\begin{table*}[t]
\caption{Health-care scenario: Mrs Smith, a 75 year old English lady, a
  former school teacher.} 
\label{tab:Healthcare}
\begin{center}
\begin{tabular}{|p{6.8cm}|p{3.8cm}|p{5.9cm}|}
\hline
Scenario & Robot skills & Cultural competence\\
\hline
\textit{The robot Aristotle detects that Mrs. Smith is in a bad mood and adopts a more cheerful voice} & Perception (Understanding facial expressions) & \\
ROBOT ARISTOTLE: How do you feel today Dorothy? & &\\
MRS DOROTHY SMITH: I feel OK but it's time for my tablets. I have diabetes. & & [Cultural Knowledge: general (4)] The UK has a pragmatic orientation.\\
A: Do you take tablets for diabetes? & Speaking (Catching key words; asking for yes/no confirmation) & [Cultural Knowledge: specific (5)] The robot is matching what Mrs. Smith says with pre-stored knowledge about her health.\\
D: Yes. &  & \\
A: Do you want me to remind you to take them? & &\\
D: Yes! I take them three times a day: morning, midday and evening. But sometimes I forget them. & &\\
A: OK. I will remind you! Please select your schedule on my screen. & Planning (Reminder) & [Cultural Knowledge: specific (6)] The robot knows that Mrs. Smith, a former school teacher, is already familiar with using a tablet. \\
\textit{The robot leans forward. Mrs. Smith selects morning, midday and evening on the screen} & Moving (Body posture), Multi-modal Interaction (Using multiple input modalities) & \\
A: Is there anything I can do for you? Can I get you some water for the tablets? & & [Cultural Knowledge: specific (7)] The robot is acquiring knowledge about what it means to Mrs. Smith to have diabetes.\\
D: Yes. That would be very nice Aristotle. & &\\ 
\textit{The robot goes to fetch a glass of water} & Planning (Retrieving an object), Perception (Locating an object), Moving (Legs, hands) &\\
\hline
\end{tabular}
\end{center}
\end{table*}

\section{Required capabilities\\ for a culturally competent robot}
\label{sec:required_capabilities}

The concepts related to cultural competence that we have defined in the
previous section are necessarily general.  In order to arrive to a set
of concrete technical requirements for cultural competent robots, we
have grounded those concepts in three tangible examples, summarized in
Tables \ref{tab:Introduction}, \ref{tab:Healthcare} and
\ref{tab:Home_and_family}.  Each example describes a possible scenario
of interaction between a culturally competent assistive robot and an
elderly person.  The examples have been written by experts in
Transcultural Nursing and draw inspiration from the rationale and
actions of culturally competent (human) caregivers.

Each table reports a pattern of sensorimotor and/or verbal interaction,
the required robot skills, as well as the cultural competence (in terms
of cultural awareness, cultural knowledge and cultural sensitivity) that
may contribute to determine the robot's behaviour. Albeit short, the
scenarios show that the following capabilities are key for a robot to
exhibit a culturally competent behaviour.

\paragraph{Cultural knowledge representation} This refers to the
  capability of storing and reasoning upon cultural knowledge, see for
  example the interaction between the robot Aristotle and Mrs.\ Smith in
  Table~\ref{tab:Healthcare}, in which the robot first uses knowledge
  (6) about Mrs.\ Smith 's work experience to tune how to introduce a
  new interaction modality (its tablet), and later acquires new
  knowledge about her habits and medical prescriptions (7).

\paragraph{Culturally-sensitive planning and execution} This refers
  to the capability to produce plans and adapt such plans depending on
  the cultural identity of the user.  Cultural sensitivity, in the
  interaction between the robot Yuko and Mrs.\ Yamada in
  Table~\ref{tab:Home_and_family}, allows the robot for planning to help
  Mrs.\ Yamada make a video call (10).

\paragraph{Culture-aware multi-modal human-robot interaction} This
  refers to the capability of adapting the way of interacting (in terms
  of gestures, choice of phrases, tone and volume of voice, etc.) to the
  user's cultural identity.  Cultural sensitivity makes the robot avoid
  asking direct questions to Mrs.\ Yamada (see
  Table~\ref{tab:Home_and_family}) and perform the proper greeting
  gestures with Mrs.\ Christou (see Table~\ref{tab:Introduction}).

\paragraph{Culture-aware human emotion and action recognition} This
  refers to the capability to interpret sensor data acquired by the
  robot during the interaction in light of cultural knowledge.  As an
  example, in Table~\ref{tab:Healthcare} the robot Aristotle correctly
  labels Mrs.\ Smith's facial expression as indicative of a bad mood,
  while in Table~\ref{tab:Home_and_family} the robot Yuko relies on
  Mrs.\ Yamada's facial expression to get feedback on its suggestion to
  make a video call.

\paragraph{Cultural identity assessment, habits and preferences
  detection} This refers to the capability to adapt general cultural
knowledge and acquire new knowledge to better fit the individual profile
of the user.  As an example, in Table~\ref{tab:Introduction} the robot
Sofia uses knowledge about the Greek culture to guess how Mrs.\ Christou
would like to be addressed (1), and uses her answer to validate its
hypothesis (2).  In Table~\ref{tab:Healthcare}, the robot Aristotle
learns Mrs.\ Smith's habits in dealing with her medical prescriptions
(5), and in Table~\ref{tab:Home_and_family} the robot Yuko brings up the
topic of video calls (8) to learn about Mrs.\ Yamada's family.

\begin{table*}[t]
\caption{Home and family scenario: Mrs Yamada, a 75 years old Japanese
  lady who performed tea ceremony in Kobe for 40 years} 
\label{tab:Home_and_family}
\begin{center}
\begin{tabular}{|p{6.8cm}|p{3.8cm}|p{5.9cm}|}
\hline
Scenario & Robot skills & Cultural competence\\
\hline
ROBOT YUKO: It is possible to test the video call with your family, if you like it. & Speaking (Avoiding direct qquestions) &\\
\textit{The robot checks for Mrs. Yamada's reaction. She smiles.} & Perception (Understanding facial expressions) &\\
MRS NAOMI YAMADA: Really? My son and daughter both live in Tokyo. My son is always busy, but he visits me during holidays. I miss my daughter so much. Her husband is Korean so she often goes to Korea. I want to call my husband, but he's now giving a lecture at school. & & [Cultural Knowledge: specific (8)] Naomi provides her personal details only when the robot brings up the topic.\\
Y: I can make a video call to your daughter, son or husband if you want. & Speaking (Catching key words and reacting) &\\
\textit{The robot checks for Mrs. Yamada's reaction} & Perception (Understanding facial expressions) &\\
N: Maybe later. I don't know how to do it. Can you give me a manual on how to do it? & & [Cultural Knowledge: general (9)] Japan is one of the most uncertainty avoiding countries on earth.\\
Y: Just tell me who you want to call, then I can help you. You are welcome to try. & Planning (Video call) & [Cultural Sensitivity (10)] Empowering: the robot is sensitive of the fact that Naomi is frightened by using unknown technology, and encourages her.\\
N: Ok, let's try. You will be my assistant! & &\\
\hline
\end{tabular}
\end{center}
\end{table*}



\section{Development methodology\\ for a culturally competent robot}
\label{sec:methodology}

The next and last step in our discussion is to propose a methodology to
develop culturally competent robots.  In order to make the discussion
more concrete, we assume that the capabilities discussed above are
organized in the functional architecture shown in
Figure~\ref{fig:architecture}. Cultural knowledge is provided offline by experts and online by users
(orange arrows) and processed to perform a cultural identity assessment
of the user.  The assessment allows for better matching to the user's
preferences the commands and parameters for the robot (blue arrows), as
well as for any smart device eventually distributed in the environment
(yellow box).

\subsection{Bootstraping cultural knowledge}
\label{Transcultural Robotic Nursing}

\begin{figure}[b]
\centering
\includegraphics[width=\columnwidth]{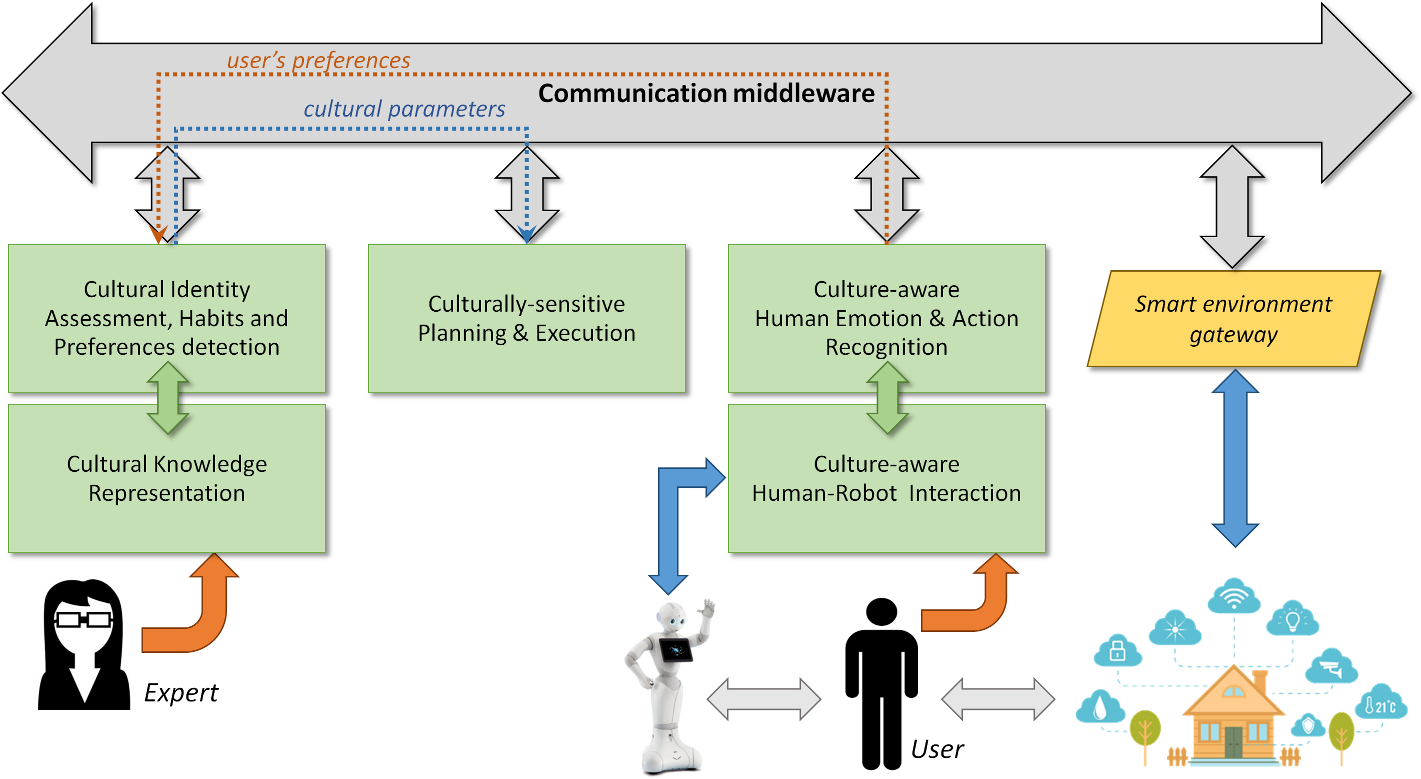}
\caption{The functional architecture of a culturally competent robot.}
\label{fig:architecture}
\end{figure}

As Figure \ref{fig:architecture} shows, the robot's attitude towards the
user is initially created based on experts knowledge from the fields of
Transcultural Nursing and Culturally Competent Healthcare.
National-level cultural information, such as Hofstede's
dimensions~\cite{Hofstede91}, complemented with specific information
about the user's cultural group, allow for making preliminary
assumptions about the expected behaviour of a cultural competent robot,
described in terms of Cultural Awareness, Cultural Knowledge and
Cultural Sensitivity~\cite{Papadopoulos06}. This preliminary assessment
can be refined on the basis of the cultural behavioural cues collected,
for example, from video recorded encounters between older people living
in sheltered housing and their caregivers, carefully analysed to avoid
stereotypic notions~\cite{Makatchev13}.

The knowledge acquired in all these steps shall ultimately produce
guidelines describing how culturally competent robots are expected to
behave in assistive scenarios. Moreover, in the perspective of a
commercial exploitation, it can allow the development of robots that are
able to autonomously acquire information and update their own knowledge
about the cultural context in which they are operating and
--–ultimately--– to re-configure their approach towards the user.

\subsection{Cultural Knowledge Representation}
\label{Cultural Knowledge Representation}

Properly encoding guidelines for cultural competence in a framework for
knowledge representation requires to take into account both
methodological and architectural aspects.

Methodological aspects include: (i) how to represent the relationship
between quantitative and qualitative knowledge about different cultural
groups; (ii) how to avoid stereotypes by allowing for differences among
individuals, while using the information about their national culture as
a hint about their cultural identity; (iii) how to automatically reason
on cultural knowledge for producing a culturally competent robotic
behaviour, i.e., plans and sensorimotor behaviours aligned with the
user's cultural identity; and (iv) how to update the knowledge base as
long as new cultural knowledge is acquired through user-robot
interaction.


Technical aspects include: which languages and tools should be used for
cultural knowledge representation,
which languages and tools should be used for querying the knowledge
base,
which reasoning tools should be adopted,
and which Application Programming Interfaces, data formats, and
protocols should be used to allow the robot to access the knowledge
base.  The field of Artificial Intelligence (AI) has a rich repertoire
of formalisms and tools for knowledge representation and
reasoning~\cite{VanHarmelenEtAl.book2008}, but how these can be applied
to the specific case of \emph{cultural} knowledge is an open and as yet
unexplored issue.


A key problem here is how to define procedures for the knowledge base
creation and updating.  In general, cultural knowledge can be defined
and introduced in the system \textit{a priori} by experts in
Transcultural Nursing or by formal and informal caregivers; or it can be
acquired at run-time through robot-user interaction; or both.  Run-time
knowledge acquisition raises the most important methodological and
technological issues, e.g., which questions should be posed to the user,
how answers should be interpreted, how the information retrieved should
be used to pose subsequent questions and to update the Knowledge Base
itself.  It also raises issues on how general cultural information known
a priori (e.g., at national level) impacts on individual
characteristics, and how the information acquired during robot-user
interaction (i.e., through explicit communication) can be merged with
the already available knowledge in order to perform a more accurate
cultural assessment.

Finally, the inclusion of personal knowledge in the cultural knowledge
base raises ethics issues in data privacy and protection, which are even
more compelling since the system may store sensitive information not
only about the users, but also about their family (e.g., names, health
status).

\subsection{Culturally-Sensitive Planning and Execution}
\label{Culturally-Sensitive Planning and Execution}

Once cultural knowledge has been explicitly produced, the challenge is
to make the robot use this knowledge to modulate its own behaviour to
match the cultural identity of the user. Technically, the ability of the
robot to ``modulate its own behavior'' translates into the ability to:
(i) form plans to achieve the robot's goals while being aware of, and
sensitive to, the user's culture; and (ii) execute the actions in these
plans in a way that is also culturally aware and sensitive.  As an
example, the three robots in Tables \ref{tab:Introduction},
\ref{tab:Healthcare} and \ref{tab:Home_and_family} may have the same
goal to help preparing the lunch, but they may achieve this goal using
different plans.  These plans may include different actions (e.g.,
Aristotle may help Mrs Smith by ordering the food online, whereas Sofia
listens to Mrs Christou chatting about cooking), or different ways to
perform an action (e.g., Yuko collaboratively prepares the lunch with
Mrs Yamada).

The field of AI has a long tradition in developing techniques for the
automatic generation and execution of action plans that achieve given
goals~\cite{Ghallab14}. Cultural aspects can contextually influence the
generation and execution of action plans in three ways:
\begin{itemize}

\item Discourage the use of certain actions; for example, to avoid
  suggesting recipes to Mrs Christou;

\item Include additional preconditions or goals, which may result in the
  inclusion of new actions; for example, with Mrs Yamada, the robot Yuko
  performs an inquiry action before committing to one action plan or
  another;

\item Induce a preference for some actions; for example, Yuko may
  encourage Mrs Yamada to cook instead of ordering food online, because
  this better complies with Mrs Yamada's need to make physical activity.

\end{itemize}

To take these influences into account, state-of-the-art approaches to
constraint-based planning~\cite{Mansouri16} shall be considered. In
addition to requirements in terms of causal preconditions (e.g., the
robot's hand must be empty to grasp an object), spatial requirements
(e.g., the robot must be in front of the user in order to interact), and
temporal constraints (e.g., the tea must be served before it gets cold),
constraint-based planning can also include constraints that pertain to
the human-robot relation, e.g., to encode the fact that the robot should
never clean a room where the user is standing: this extension of
constraint-based planning is particularly suited to generate plans that
take into account cultural constraints and, in general, ``human-aware
planning'' \cite{Kockemann14}. 

\subsection{Culture-aware multi-modal Human-Robot Interaction}
\label{Culture-aware interaction}

Once a proper course of actions (including both motion and speech) has
been planned taking into account the user's cultural identity, actions
must be executed and feedback must be considered to monitor their
execution. In this context, Human-Robot Interaction plays a crucial role
in enabling the robot with cultural competence. On the one hand, the way
the robot behaves and speaks can produce different impacts and
subjective experiences on the user; on the other hand, what the user
says and does is the key for the robot to acquire new knowledge about
the user, and consequently refine and improve its cultural competence.
%
%
As a prerequisite, the robot shall be equipped with motor capabilities
that are sophisticated enough to allow it to exhibit its cultural
competence through motions, gestures, posture, speech; similarly, it is
mandatory that the robot (and possibly  the  environment) is equipped
with sensors and devices for multimodal audio / video / haptic
interaction that allow providing feedback to the modules for planning,
action execution and monitoring, as well as perceiving the nuances of
human behaviour in different cultures. Moreover, communication devices
allowing for a simplified interaction may also be fundamental for frail
older adults.

The role played by robot-user \textit{verbal communication} shall be
carefully considered, as it is the primary way of interaction, possibly
allowing to acquire new knowledge and update the Cultural Knowledge
Base. We argue that, due to the current limitation in natural language
understanding, semantic comprehension shall be limited to the
recognition of relevant keywords, that the robot will use to react
accordingly, by asking a confirmation through a simple multiple choice
(e.g., yes/no) question. Additional touchscreen-based interfaces (either
embedded on the robot or carried by users, e.g., tablets and
smartphones) might be used to complement the verbal interaction
modality.

\subsection{Culture-aware Human Emotion and Action Recognition}
\label{Culture-aware Emotion and Action}

The robot's perceptual capabilities shall include the ability to
estimate human \textit{emotions} (joy, sadness, anger, surprise) and
recognize human \textit{actions}. If the robot operates in a smart ICT
environment, the usage of lightweight wearable sensors that do not
interfere with daily activities shall be explored (e.g., smartwatches or
sewable sensors).
%

Lastly, the robot shall be equipped with a module to detect and
recognize \textit{daily activities}, i.e., combinations of primitive
actions performed in different contexts and places of the house (e.g.,
walking, cleaning, sitting on a sofa, etc.) \cite{Bruno14}.

\subsection{Cultural Identity Assessment, Habits and Preferences detection}
\label{Cultural Identity Assessment}

As time progresses and the robot has more and more interactions with the
user, daily activities and manners (a subset of social norms that
regulate the actions performed by the user towards other humans, or even
the robot itself) may be assessed to determine the long-term
\textit{habits} of the human companion.  Verbal interaction, as well as
the assessment of user's, emotions, actions, daily activities, manners,
and habits, will ultimately provide an input to perform a cultural
assessment of the user, updating the knowledge that the robot has about
the user's cultural identity.

The aforementioned capabilities shall involve procedures to merge and
interpret sensor data acquired by the robot and by the smart ICT
environment at the light of cultural knowledge that is already stored in
the system.  Indeed, cultural knowledge can play a fundamental role at
all levels of perception, ranging from basic object recognition to the
detecting of daily activities, manners and habits.  For instance, if the
system is uncertain if a purple object in the fridge is a slice of pig
liver or an eggplant, cultural information about the alimentary customs
of the users (that maybe are vegetarians) could help to disambiguate.

\section{Testing and Evaluation}
\label{Testing and Evaluation}

Once the technology for a cultural competent robot are in place, its
impact on the target user group shall be empirically evaluated.  A
typicaly protocol would divide participants in an experimental arm,
interacting with robots with cultural customization, and a control arm,
interacting with robots without cultural customization.  Beside the
cultural customization, the two arms should be use setups that as as
similar as possible.

In the case of evaluation with elderly participants, an ethically
sensitive and detailed protocol that describes the screening,
recruitment, testing and analytical procedures must be produced and
scrutinised by relevant ethics committees. Testing should involve older
adults belonging to different cultural groups, who possess sufficient
cognitive competence to participate and who are assessed as sufficiently
unlikely to express aggression during the testing period, together with
nominated key informal caregivers (e.g. close family members).
%
%
%

End-user evaluation shall be aimed at evaluating the capability of
culturally competent systems to be more sensitive to the user's needs,
customs and lifestyle, thus impacting on the quality of life of users
and their caregivers, reducing caregiver burden, and improving the
system's efficiency and effectiveness. 
%
%
Quantitative outcomes of interest and measurement tools shall include
the following (pre and post testing).

\setcounter{paragraph}{0}

\paragraph{Client perception of the robot's cultural competence} 
   Measurement tool: Adapted RCTSH Cultural Competence Assessment Tool
   (CCATool) \cite{Papadopoulos04}. The tool measures clients'
   perceptions of the robot's cultural awareness, cultural safety,
   cultural competence and cultural incompetence, and includes items
   associated with dignity, privacy and acceptability.

\paragraph{Client and informal caregiver health related quality of life} 
  Measurement tool: Short Form (36) Health Survey (SF-36)
  \cite{Hays93}. The SF-36v2 is a multi-purpose, short-form health
  survey proven to be useful in surveys of general and specific
  populations, including older adults. It measures general health,
  bodily pain, emotional role limitation, physical role limitation,
  mental health, vitality, physical functioning and social
  functioning. Each dimension score has values between 0 and 100, in
  which 0 means dead and 100 perfect health.

\paragraph{Informal caregiver burden} 
  Measurement tool: The Zarit Burden Inventory (ZBI) \cite{Zarit80}. The
  ZBI is a widely used 22-item self-report inventory that measures
  subjective care burden among informal caregivers. Its validity and
  reliability have been widely established. The scale items examine
  burden associated with functional/behavioural impairments and care
  situations. Each item is scored on a 5-point Likert Scale, with higher
  scores indicating higher care burden among informal caregivers.

\paragraph{Client satisfaction with the robot}
  Measurement tool: Questionnaire for User Interface Satisfaction (QUIS)
  \cite{Chin88}. This scale evaluates whether the clients are satisfied
  with the interaction process including its efficiency and
  effectiveness. It should be adapted so that ``the software'' is
  replaced by ``the robot''.

Clients and their informal caregivers shall also be invited to
participate in qualitative interviews to elicit discussions about their
perceptions of the robot's cultural competence, quality of service
provided, impact upon independence and autonomy, as well as --- very
importantly --- experiences related to configuring the system by
injecting cultural knowledge before operations.

\section{Discussion and Conclusions}
\label{sec:discussion}
\addtolength{\textheight}{-45mm}

This position paper has discussed foundations, rationale and a possible
methodology for developing and evaluating a culturally competent robot,
i.e., a robot able to autonomously re-configure its way of acting and
speaking, when offering a service, to match the culture, customs, and
etiquette of the person it is assisting.  We believe that cultural
competence is a necessary, although so far understudied, ingredient for
any social, personal or, and especially, assistive robot.

The methodology proposed in this article is being implemented in the
project CARESSES which is, to the best of our knowledge, the first
attempt to build culturally competent robots. CARESSES' starting
technology includes the humanoid robot \textit{Pepper}, produced by
Softbank Robotics Europe, as well as the \textit{iHouse}, a
Japanese-based duplex apartment fully embedded with sensors and
actuators for home automation developed by the Japan Advanced Institute
of Science and Technology.

In its current stage of development, CARESSES has gone through an
initial investigation phase aimed at producing guidelines for
Transcultural Robotic Nursing, as described in Section
\ref{Transcultural Robotic Nursing}, and followed by a development phase
aimed at designing and implementing components that realize the key
technical capabilities discussed in Sections~\ref{Cultural Knowledge
  Representation}, \ref{Culturally-Sensitive Planning and Execution},
\ref{Culture-aware interaction}, \ref{Culture-aware Emotion and Action}
and \ref{Cultural Identity Assessment} above.  These components will be
integrated in universAAL~\cite{Ferro15}, a software platform for open
distributed systems of systems that resulted from a consolidation
process conducted within an EU Project.  Testing and end-user evaluation
will follow the procedure in Section~\ref{Testing and Evaluation} and
include at least: ten clients who primarily identify themselves with the
white-English culture, ten clients who primarily identify themselves
with the Indian culture, and ten clients who primarily identify
themselves with the Japanese culture.
Each client will adopt a Pepper robot for a total of 18 hours over a
period of two weeks, which should allow for enough time for a culturally
customized Pepper robot to acquire knowledge about the individual
cultural characteristics of the assisted person and provide culturally
competent interactions and service, which will then be evaluated through
quantitative tools and qualitative interviews.


CARESSES only aims at producing a first prototype, which shall be
further evaluated and refined before drawing definitive conclusions
about the impact of cultural competence in assistive robotics.
Nonetheless, we strongly believe that the CARESSES pilot will be a
foundational breakthrough in culturally competent robotics, and it will
be invaluable in paving the way for future similar studies.

\section*{Acknowledgement}

This work has been supported by the European Commission Horizon2020
Research and Innovation Programme under grant agreement No. 737858, and
from the Ministry of Internal Affairs and Communication of Japan.  We
thank all the members of the CARESSES consortium, listed in the project
homepage \url{www.caressesrobot.org}.

\bibliography{bibliografia}

\end{document}